# SIZES OF VOIDS AS A TEST FOR DARK MATTER MODELS

Sebastiano Ghigna[1], Stefano Borgani[2],

Silvio A. Bonometto[1], Luigi Guzzo[3],

Anatoly Klypin [4], Joel R. Primack[5],

Riccardo Giovanelli & Martha P. Haynes[6]

[1] Dipartimento di Fisica dell'Università di Milano Via Celoria 16, I-20133 Milano, Italy *and* INFN, Sezione di Milano

[2] INFN, Sezione di Perugia, c/o Dipartimento di Fisica dell'Università Via A. Pascoli I-06100 Perugia, Italy *and* SISSA, International School for Advanced Studies, Via Beirut 2-4, I-34014 Grignano, Trieste, Italy

[3] Osservatorio Astronomico di Brera, Sede di Merate, Via Bianchi 46, I-22055 Merate (CO), Italy

[4] Astronomy Department, New Mexico State University Box 30001 – Dept. 4500, Las Cruces, NM 88003-0001, USA *and* Astro-Space Center, Lebedev Phys. Inst., Moscow, Russia

[5] Institute for Particle Physics, University of California, Santa Cruz, CA 95064, USA

[6] Department of Astronomy and National Astronomy and Ionosphere Center (The National Astronomy and Ionosphere Center is operated by Cornell University under a cooperative agreement with the National Science Foundation) Space Sciences Building, Cornell University, Ithaca, NY 14853, USA

**Abstract**

We use the void probability function (VPF) to study the distribution of galaxies in a volume–limited subsample of the Perseus–Pisces survey. We compare observational results with theoretical predictions based on high–resolution N–body simulations for two realizations of the Cold+Hot Dark Matter (CHDM) model and for unbiased ($b = 1$) and biased ($b = 1.5$) CDM models in a 50 $h^{-1}$Mpc box. We identify galaxies as peaks of the evolved density field. Overmerged structures are fragmented into individual galaxies so as to reproduce both the correct luminosity function (after assuming $M/L$ values for the resulting galaxy groups) and the two–point correlation function. We also try to reproduce the observational biases of the observational data as best as we can. Our main result is that on intermediate $2 - 8\, h^{-1}$Mpc scales the void–probability function for the standard CHDM model with $\Omega_{cold}/\Omega_{hot}/\Omega_{bar} = 0.6/0.3/0.1$ exceeds the observational VPF with a high confidence level. CDM models produce smaller VPF, whose shape is independent of the biasing parameter. We verify the robustness of this result against changing the observer position in the simulations and the threshold for galaxy identification.

# 1. Introduction

Large voids were the unexpected finding when galaxy redshift samples were first compiled. The void probability function (VPF) was applied to provide a quantitative estimate of their probability and provides information about the large–scale galaxy distribution which goes beyond the two-point correlation analysis. In this *Letter*, we will use the VPF in order to compare voids in a volume–limited subsample of the Perseus–Pisces Survey (PPS, Giovanelli & Haynes 1991) with results of N–body simulations based on CDM and CHDM models (see, e.g., Klypin et al. 1993 and references therein)

Several attempts have been pursued to analyze the void statistics in observational data, like the SSRS survey (Maurogordato, Schaeffer, & da Costa 1992) and the 1.2 Jy IRAS redshift survey (Bouchet et al. 1993), and the CfA survey (Vogeley et al. 1994). Fry et al. (1989) did a void analysis for a preliminary version of the PPS and compared the results to N–body simulations. Weinberg & Cole (1992) used the VPF to discriminate between Gaussian and non–Gaussian initial conditions in N–body simulations. A crucial point when using VPF to test models of structure formation concerns the identification of galaxies, since changing the efficiency of galaxy formation in underdense regions has a dramatic impact on the resulting VPF (Betancort–Rijo 1990; Einasto et al. 1991; Little & Weinberg 1994). Thanks to the high mass resolution of simulations used here, we identify galaxies as high peaks of the evolved density field. However, even with this choice, there are many ways to assign galaxies to peaks.

It is well known that the mass function of dark matter halos has a different shape as compared with the observed luminosity function of galaxies. A number of effects contribute to the difference. Lack of numerical resolution results in formation of "overmergers" – large halos in central parts of groups and clusters (e.g., Gelb & Bertschinger 1993). Those halos in reality should correspond to a few galaxies, not one. Besides the numerical effects, which mainly affect the high-mass tail of the mass function, there are uncertainties related to largely unknown effects of the feedback from star formation. It is very likely that if we take them into account, the assumption of constant mass-to-light ratio might not be valid anymore, with small halos producing much less stars and light as compared to naive mass-follow-light predictions. In any case, when making comparisons with observations, one needs to take different halos with different "weight". We found that it is easier to use the language of halo splitting with a constant mass-to-light ratio as a substitute for weights of halos. It also gives us a rough estimate of the mass-to-light ratio, which would be typical for the model.

## 2. Observational and simulated catalogs.

We consider a volume–limited subsample (VLS) of the PPS survey with the limiting magnitude $M_{\rm lim} = -19 + 5\log h$, corresponding to $79\,h^{-1}$Mpc for the limiting depth. To avoid high galactic extinction, we take the sample boundaries $22^h \leq \alpha \leq 3^h\,10^m$, $0° \leq \delta \leq 42°\,30'$, which include 1032 galaxies. The resulting mean galaxy separation is $d = 5.2\,h^{-1}$ Mpc. The PPS survey consists mainly of highly accurate 21-cm HI line redshifts, partly unpublished, obtained with the NAIC 305-m telescope in Arecibo and with the NRAO 92-m telescope formerly in Green Bank. The radio data are complemented with optical observations of early-type galaxies carried out at the 2.4-m McGraw-Hill telescope, so that the redshift survey is virtually 100% complete for all morphological types (for details see e.g. Giovanelli & Haynes 1989; 1991).

Observational data are compared to a set of Particle-Mesh simulations, which were run for $50\,h^{-1}$ Mpc boxes on a $512^3$ grid. Four simulations are considered: two realizations of COBE–compatible Cold+Hot Dark Matter model (CHDM$_1$ and CHDM$_2$) with $\Omega_{cold}/\Omega_{hot}/\Omega_{bar} = 0.6/0.3/0.1$ and a realization of Cold Dark Matter model with bias factors $b = 1$ and 1.5 (CDM1 and CDM1.5). The latter used the same random numbers as CHDM$_1$. As discussed elsewhere (Nolthenius, Klypin, & Primack 1994; Bonometto et al. 1994), the largest waves in the CHDM$_1$ and CDM simulations have amplitude about 1.3–1.4 times larger than expected for an average realization. We estimate that this statistical fluke, causing extra large–scale power, would be generated once in ten or twenty realizations.

Here we give more details of our galaxy identification scheme (for further details, see Ghigna et al. 1994). We start with assuming a suitable mass–to–light ratio $(M/L)$ (which will be varied) and calculate the expected total luminosity $L_{\rm total}$ and the total mass $M_{\rm total} = M/L \times L_{\rm total}$ of galaxies in the computational volume $V = (50\,h^{-1}{\rm Mpc})^3$.

In order to do that we take the following luminocity function of real galaxies: $\phi(L) = \phi_* (L/L_*)^\alpha \exp(-L/L_*)$, where $\phi_* = 1.56 \times 10^{-2} h^3 \mathrm{Mpc}^{-3}$, $\alpha = -1.07$ and $L_*$ corresponds to the absolute magnitude $M^* = -19.68$ (Efstathiou, Ellis & Peterson 1988). We integrate this luminosity function over all luminosities above $L_\mathrm{lim}$ corresponding to the magnitude limit $M_\mathrm{lim}$ of the observational sample. Then we construct a list of "galaxies". The number of galaxies in the list is the number of real galaxies expected in our computational volume: $N_\mathrm{gal} = V/d^3$, where $d$ is the mean distance between galaxies in the PPS sample. For each "galaxy" we assign a luminosity in such a way that the sum of all luminocities is $L_\mathrm{total}$ and their distribution is the observed luminosity function $\phi(L)$. Each "galaxy" will correspond to a halo or to a part of a halo.

Now we find dark halos in the simulations. Each halo corresponds to a peak of density on our $512^3$ mesh. Only the most massive $N_\mathrm{halos}$ halos will be include in the final "catalog". The mass assigned to a halo is the mass inside a cube of $3 \times 3 \times 3$ computational cells centered on the maximum. This effectively corresponds to halo radius $R_h \simeq 182\, h^{-1}\mathrm{kpc}$. We select the most massive halos, so that the total mass in them is equal to the expected mass in observed galaxies $M_\mathrm{total}$. This gives $N_\mathrm{halos} = 700 - 900$ with overdensities larger than 500 for CHDM simulations and larger than 1100 for CDM.

Finally, "galaxies" in the first list are distributed among DM halos in the second list. We take the most massive galaxy and assign it to the most massive halo in the simulation. Because the halo is more massive than the galaxy, there is some halo mass left for another galaxy. If the mass left in the halo is larger than the mass of the second largest galaxy, we "assign" the galaxy to the halo. If not, then the next most massive galaxy is tried and so on until we find a galaxy, the mass of which is smaller than the remaining mass of the halo. The procedure is repeated untill all mass of the first halo is split between galaxies. Then we take the second most massive halo and work with it. Note that most massive galaxies will be placed in most massive halos.

In Figure 1 we plot the ratio of the number density of selected halos, $n_\mathrm{halos} = N_\mathrm{halos}/V$ to that of galaxies, $n_\mathrm{gal} = d^{-3}$, as a function of assumed $M/L$. Two features are apparent: (i) $n_\mathrm{halos}$ sensitively depends on $M/L$; (ii) once $n_\mathrm{halos}$ is given, the corresponding $M/L$ sensitively depends on the model and on the choice of the biasing parameter, but not on the realization (note the strong similarity between the CHDM$_1$ and CHDM$_2$ curves). An upper bound on $M/L$ is provided by the condition that $n_{\text{halos}}$ does not exceed $n_{\text{gal}}$ (no fragmentation of halos).

Since after the fragmentation more then one galaxy can be associated with a halo, the galaxies should be given different velocities. Let **v** be the velocity of a halo; then the velocity $\mathbf{v}_{\text{gal}}$ of a galaxy belonging to that halo is $\mathbf{v}_{\text{gal}} = \mathbf{v} + \Delta \mathbf{v}_{\text{gal}}$, where $\Delta \mathbf{v}_{\text{gal}}$ results from local virial motions. We assume that $\Delta \mathbf{v}_{\text{gal}}$ has Gaussianly distributed components, with variance $\langle \Delta v_{\text{gal}}^2 \rangle / 3$. We take $\langle \Delta v_{\text{gal}}^2 \rangle = GM/R_h$. We verified that adding such virial velocities does not significantly alter the small–scale pairwise galaxy velocity dispersion. Once a suitable $\mathbf{v}_{\text{gal}}$ is assigned to each galaxy, we *observe* the galaxy distribution by randomly placing 20 "observers" The greatest depth obtainable from the $50 h^{-1}$Mpc box is $86.6 h^{-1}$Mpc. This exceeds the $79 h^{-1}$Mpc depth of the observational volume limited sample only within a small angle around the diagonal of the box. One therefore has to consider replicae of the simulation, using the periodic boundary conditions. Multiple use occurs for $\sim 20$ % of the box and the line of sight is always differently directed in those zones which are used more than once. Only samples containing the same number of galaxies as in the real PPS sample (with a tolerance of 30 objects) were accepted.

In order to fix $M/L$ for the halo fragmentation, we impose that the resulting galaxy distribution has a 2–point correlation function, $\xi(r)$, close to that observed for PPS. For CHDM$_1$ the anomalous amount of power generates an exceedingly large $\xi(r)$ at all scales, even allowing for no halo fragmentation ($M/L = 275 h$; Ghigna et al. 1994), which corresponds to the lowest allowed density threshold (the correlation length was $\sim 10\, h^{-1}$Mpc ). As for the other models, reliable fits to observational data are attained for $M/L = 250 h$, $500 h$ and $600 h$ for CHDM$_2$, CDM1.5 and CDM1, respectively. These $M/L$ values, which we will use in the following VPF analysis, are consistent with those suggested by observations of galaxy groups (e.g., Ramella, Geller, & Huchra 1989; Nolthenius 1993; Moore, Frenk, & White 1993; Mamon 1993), although within the quite large uncertainties.

**3. The VPF analysis**

In order to estimate the VPF, we sample the galaxy distribution with random spheres of different radii $R$. Centers are taken to be at distance $> R$ from sample boundaries. We take $N_R = 2\, V_T/V_R$ of such spheres ($V_T \simeq 1.5 \times 10^5 h^{-3} {\rm Mpc}^3$ is the sample volume, $V_R = 4\pi R^3/3$), where the factor 2 accounts for the presence of clustering (e.g., Fry & Gatzañaga 1994).

In Figure 2 we plot the resulting VPF for both observational (solid curves) and simulated (dotted curves) galaxy samples. Also plotted as a dashed curve is the Poissonian VPF. For the observational data, the plotted error bars are $3\sigma$ uncertainties estimated from 20 bootstrap resamplings. For each model we plot results for five typical observers, although we realized the analysis taking 20 observers. The scatter among observers, which is a measure of the *sky–variance*, is always small and of the same order of the bootstrap errors. In Table 1 we report the VPF values at five different scales, along with the *sky–variance* for the simulations). As a major result, it turns out that both CHDM simulations produce too many voids of $2-5\,h^{-1}{\rm Mpc}$ size. The discrepancy with respect to the observational data is large for CHDM$_1$, which is known to have an anomalous amount of large–scale power, while it is less pronounced for the more typical CHDM$_2$ realization. Both CDM models correctly reproduce the PPS data. This agrees with the claim of Little & Weinberg (1994) that, within a fixed biasing scheme, the VPF does not depend sensibly on the spectrum normalization. However, the CDM simulations have the same excess power as CHDM$_1$ and a typical CDM realization might generate a VPF smaller than the real sample does.

In order to verify the robustness of this result, we checked that the VPF is independent of the choice of the observer. Instead of randomly choosing the observers, we attempted to pick them out so as to reproduce the main features of the PPS galaxy distribution (i.e., the presence of a nearby void region and of a Persues-Pisces–like supercluster at $\simeq 50\, h^{-1}$ Mpc). We find that such observers measure a VPF which is always similar to that seen by a random observer.

A further possible way to reconcile the void structure of CHDM with observations could be lowering the threshold for galaxy identification, so as to generate a more space–filling distribution. In order to be conservative, we tried a threshold below the limiting one fixed by the no halo–fragmentation scheme. In Table 1 we report the VPF values obtained by taking twice as many halos as the expected galaxies and then randomly selecting half of such halos, each associated with a single galaxy. The results for simulations are based on 20 observers for each model. This suggests that the VPF behaviour of simulated data sets is caused by the model rather than by how we pick galaxies out of the evolved density field.

As emphasized by Kauffmann & Melott (1992), a potential problem in the VPF analysis of N–body simulations resides in finite–volume effects, which reduce the number of voids of larger than one fourth of the box size. We expect this effect to be relevant at scales ($> 6\,h^{-1}$Mpc) where the VPF drops to very small values. However, it should not seriously affect the results at smaller scales ($< 3\,h^{-1}$Mpc), where our VPF analysis already discriminates between DM models (see Figure 1 and Table 1).

## 4. Conclusions

In this Letter we used the void probability function as a discriminator between dark matter models. We compared results for a volume limited sample of the Persues–Pisces Survey (PPS) to those obtained from artificial data sets extracted from high–resolution N–body simulations for the biased and unbiased CDM models and the CHDM model with 30% of neutrinos. In order to perform a realistic comparison, we attempted to reproduce the observational setup in the construction of the simulated samples.

We tune parameters of our galaxy finding algorithm so as to reproduce both the observed galaxy luminosity function and two–point correlation function. We find that, on the scale of our resolution ($\sim 200$ kpc), the mass-to light ratio, which we need to assign to dark matter halos in the CHDM model ($M/L \sim 250\,h$) is reasonably consistent with observational results for groups of galaxies. The CDM model predicts higher $M/L$ values ($\sim 500\,h$).

Our main results can be summarized as follows.

**a)** The CHDM model tends to overproduce voids of $2-8\,h^{-1}$Mpc size. Taking 20 different observers in each simulation box, none of these measures a VPF as low as that for PPS. The VPF for the CHDM model is inconsistent with the observational VPF at more than $3\sigma$ level for all scales.

**b)** Both unbiased and biased CDM models fare better on these scales. They produce a VPF close to the observed one. This seems to indicate that the void statistics is more sensible to the shape of the linear spectrum and the nature of the dark matter than to the amount of bias. However, our CDM realization is based on the same random numbers as $CHDM_1$, which had a statistical fluke on long waves. A comparison between the VPF's of $CHDM_1$ and $CHDM_2$ (the later has more typical initial realization) shows that the CDM model might give VPF's smaller than those measured on the simulations available here. Additional simulations will be needed to check that.

**c)** The above results are robust: they depend neither on the observer location, nor on the threshold for the galaxy identification. Choosing observers so to have a foreground void region and a supercluster at $\sim 50\,h^{-1}$ Mpc does not significantly change the VPF estimate. We also verify that lowering the threshold for galaxy identification does not reconcile the CHDM model with observational data. Further tests with different galaxy identification procedures would be desirable, however.

The above results seem rather surprising in light of the remarkable ability of the CHDM model to pass a series of observational constraints over a rather large scale–range (e.g., Klypin et al. 1993 and references therein). It appears that the VPF provides useful constraints on the composition of the cold, hot, and baryonic mix. In fact, our results indicate that the void statistics crucially depend on the ratio of power of initial fluctuations at galactic ($\sim 1$ Mpc) and intermediate ($\sim 10$ Mpc) scales. The composition of the dark matter may also be important, as the dark matter in the voids has an enhanced hot/cold ratio in CHDM, which helps suppress galaxy formation there. Recently, some of us (Klypin et al. 1994) compared the predictions of CHDM models with $\Omega_{cold}/\Omega_{hot}/\Omega_{bar} = 0.6/0.3/0.1$ and $0.675/0.25/0.075$ with recent observational data on the abundance of damped Ly$\alpha$ systems up to redshift $\gtrsim 3$. The increase of small–scale power in the latter model improves the performance of CHDM on this test. New simulations will allow us to see whether the VPF also agrees better.

**Acknowledgements.** JRP acknowledges support from the NSF. AK and JRP utilized the CONVEX C3880 at the National Center for Supercomputing Applications, University of Illinois at Urbana–Champaign. SB thanks NMSU and UCSC for their hospitality during the first phase of preparation of this work. SG thanks Fabio Governato for discussions.

# Figure Captions

**Figure 1.** The ratio $n_{\rm halos}/n_{\rm gal}$, where $n_{\rm halos}$ is the number density of halos selected to yield galaxies with expected density $n_{\rm gal}$, is shown at varying mass–to–light ratio $hM/L$ on $\sim 200$ kpc scale typical for galaxy groups. The continuous and (practically identical) short–dashed curve are for the two runs of the Cold+Hot Dark Matter Model (CHDM$_1$ and CHDM$_2$, respectively). The long dashed curve is for Cold Dark Matter with bias $b = 1.5$ (CDM1.5) and the short–dashed one is for $b = 1.0$ CDM (CDM1). Only values with $n_{\rm halos}/n_{\rm gal} \leq 1$ are meaningful.

**Figure 2.** The scale–dependance of the Void Probability Function $P_0(R)$ is shown for the $M_{\rm lim} < 19$ volume-limited sample (VLS) of the Perseus–Pisces Survey (continuous curve) and for five different artificial VLS's obtained from each simulation (dotted curves). The five realizations of artificial VLS's have different observer positions but the same number of galaxies as in the real VLS. The dashed curve is what one expects for a Poissonian distribution.

Table 1: The VPF at various scales $R$ for the volume–limited sample (VLS) of the Perseus–Pisces Survey (PPS) and for the artificial VLS's built up from the simulations. At each scale the first line refers to artificial galaxy samples, where the galaxy finding algorithm described in the text is exploited. The second line illustrates the stability of the results. No halo fragmentation was used in this case. Twice more numerous subsample of halos was chosen, of which only a random half fraction was kept for the final analysis. Errors are three standard deviations over 20 bootstrap resamplings and over 20 different observers for real and simulated samples, respectively.

| $R$ ($h^{-1}$Mpc) | | | $P_0(R) \cdot 10^2$ | | |
|---|---|---|---|---|---|
| | PPS | CHDM$_1$ | CHDM$_2$ | CDM1.5 | CDM1 |
| 2.1 | $84.6 \pm 0.6$ | $88.7 \pm 4.5$ | $87.7 \pm 2.1$ | $84.5 \pm 2.1$ | $84.1 \pm 2.1$ |
| | | $88.1 \pm 2.7$ | $87.1 \pm 1.8$ | $83.2 \pm 2.4$ | $82.4 \pm 2.7$ |
| 3.1 | $64.8 \pm 2.7$ | $76.5 \pm 4.2$ | $72.8 \pm 6.0$ | $65.2 \pm 5.4$ | $64.1 \pm 6.0$ |
| | | $74.5 \pm 8.7$ | $71.7 \pm 5.4$ | $61.9 \pm 5.4$ | $60.8 \pm 6.9$ |
| 4.7 | $31.1 \pm 5.7$ | $56.9 \pm 9.0$ | $48.7 \pm 9.3$ | $37.3 \pm 6.9$ | $35.4 \pm 9.6$ |
| | | $55.2 \pm 15.3$ | $46.7 \pm 12.0$ | $32.6 \pm 12.3$ | $31.5 \pm 12.0$ |

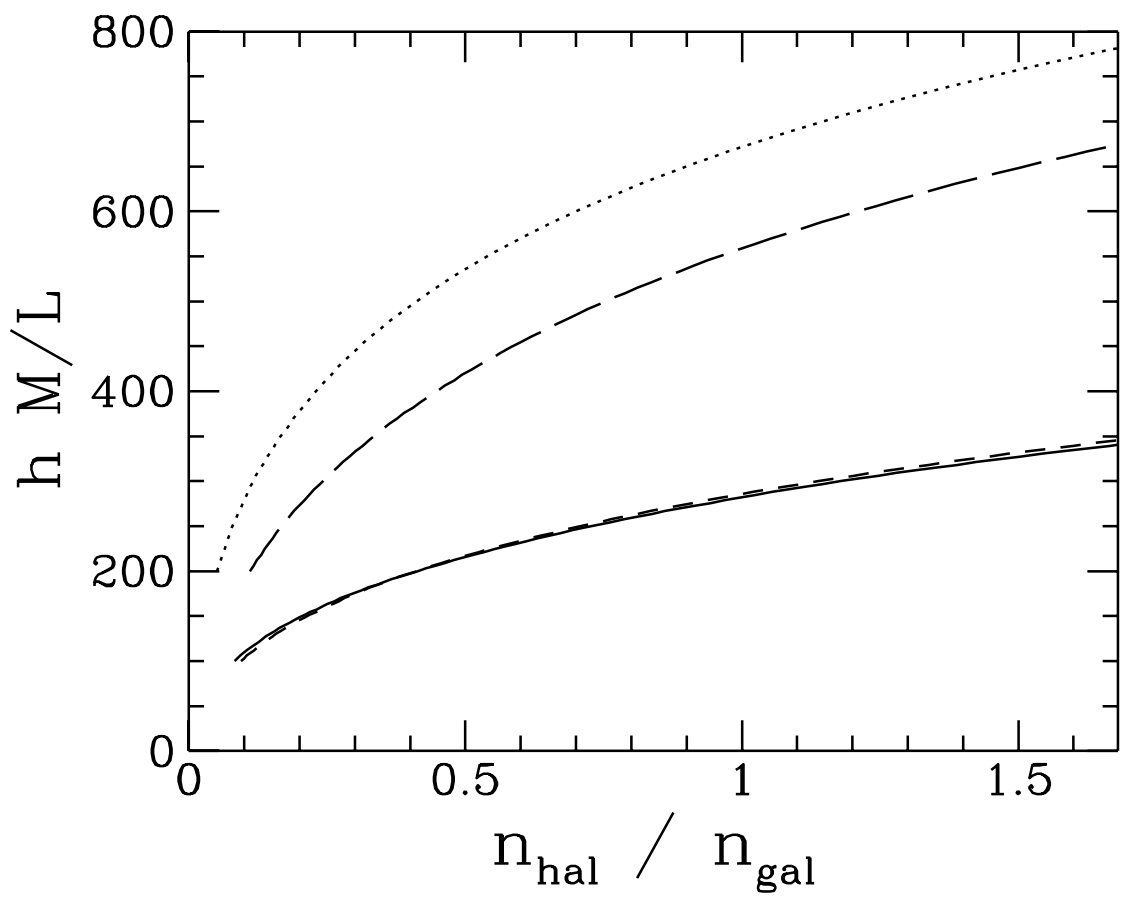

Figure 1

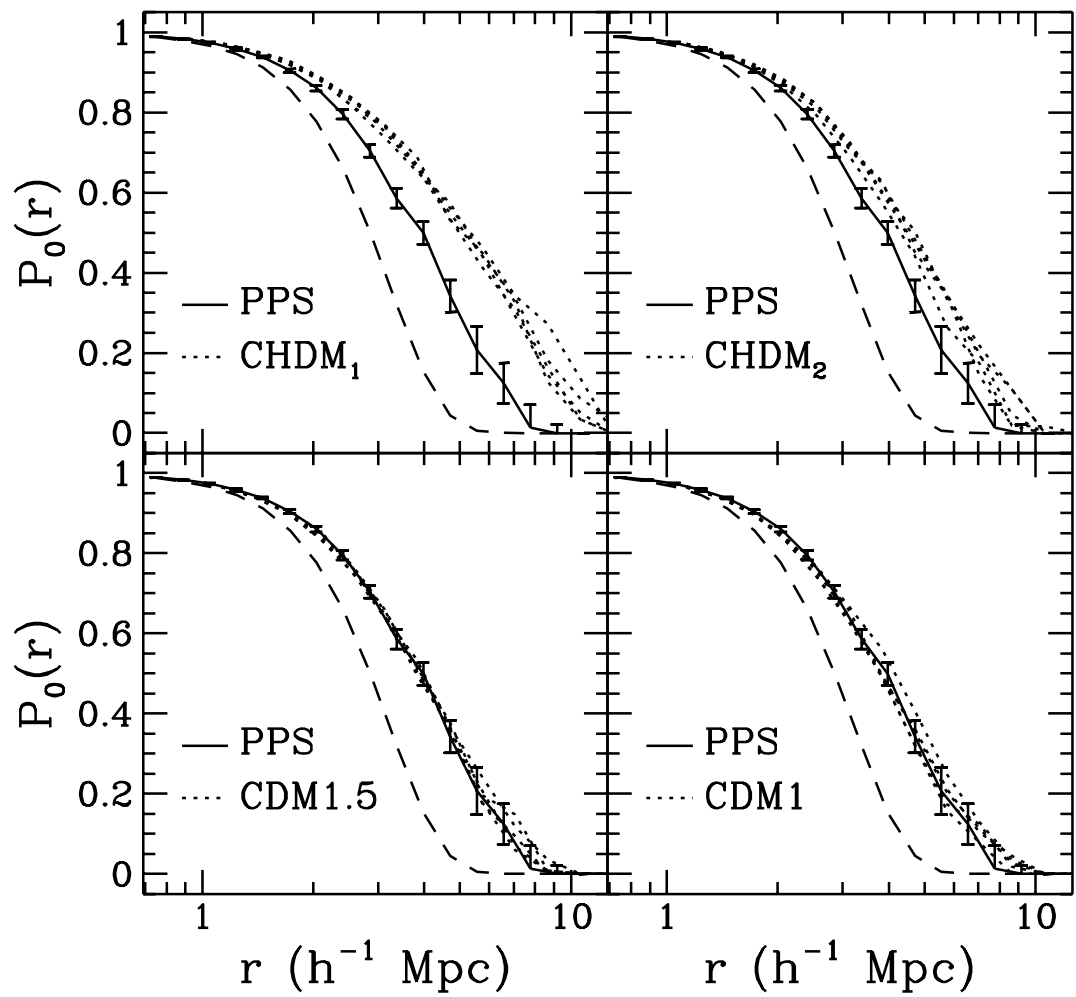

Figure 2